\documentclass[myclassdoc,rjpdebug]{rjparticle}

\usepackage[utf8]{inputenc} 
\usepackage[T1]{fontenc}

\usepackage{graphicx}
\usepackage[labelformat=simple]{subcaption}

\usepackage{soul}
\usepackage{multirow}

\title{Waiting times for sea level variations in the Port of Trieste: \\ a computational data-driven study} 

\author[1,$a$]{Gabriel Tiberiu Pan\u{a}}

\author[1,$b$]{Paul-Adrian Gog\^{i}\c{t}\u{a}}

\author[2,$c$]{Alexandru Nicolin-\.{Z}aczek}

\affil[1]{Faculty of Physics, University of Bucharest, Atomi\c{s}tilor 405, M\u{a}gurele, Romania\\
\email[a]{gabriel.pana@s.unibuc.ro}, \email[b]{paul.gogita@s.unibuc.ro}}

\affil[2]{Institute of Space Science, Atomi\c{s}tilor 409, M\u{a}gurele, Romania\\
\emailca{c}{alexandru.nicolin@spacescience.ro}}


\columntitle{Waiting times for sea level variations}

\begin{document}
\nolinenumbers
\maketitle
\begin{abstract}
We report here a series of detailed statistical analyses on the sea level variations in the Port of Trieste using one of the largest existing data catalogues that covers more than a century of measurements. We show that the distribution of waiting times, which are defined here akin to econophysics, namely the series of shortest time spans between a given sea level $L$ and the next sea level of at least $L + \delta$ in the catalogue, exhibits a distinct scale-free character for small values of $\delta$. For large values of  $\delta$, the shape of the distribution depends largely on how one treats the periodic components embedded in the sea level dataset. We show that direct analyses of the raw dataset yield distributions similar to the exponential distribution, while pre-processing the sea level data by means of a local averaging numerical recipe leads to Pareto-Tsallis distributions. 
\end{abstract}

\section{Introduction}
\label{sec:introduction}
The unprecedented availability of observational data, coming both from historical sources, which are now digitized, as well as automatized natively digital systems, allows for unparalleled investigations into the world of complex systems. The topical coverage of complex systems is immense and a rapid literature survey will show that domains as diverse as linguistics~\cite{piantadosi2014}, somnology~\cite{lo2013, lima2017}, magnetism~\cite{desousa2020} and acoustics~\cite{desousa2019}, to name just a few, are now commonly studied to show the intrinsic similarity exhibited by systems which are governed by distinct underlying mechanisms. From a different perspective, one could say that having such unparalleled access to both structured and unstructured data has cross-fertilized seemingly unrelated fields to an extent hard to imagine a few decades back and has helped us \textit{move away from the what towards the why}~\cite{kar2020}. 

Many tools have been proposed and used to assess the dynamics of complex systems, but the so-called distribution of waiting times acquired a distinct position. The concept was initially introduced for financial markets as {\it investment horizon}, {\it i.e.}, the smallest time interval needed for an index to vary by a given amount, and -- to give only two examples -- it was successfully used to describe optimal investment strategies~\cite{simonsen2002} and the gain-loss asymmetry for real and artificial stock indices~\cite{siven2009}. This definition was adapted in the context of seismic studies to investigate by computational means the distributions of time intervals between earthquakes bearing specific properties. To this end, the waiting time was defined as the shortest time interval needed to find an earthquake of magnitude of at least $M+\delta$, with $\delta$ a given constant (or threshold), after an earthquake of magnitude $M$ was observed. Remarkably, we have shown using a series of open-source earthquake analysis tools~\cite{pana2021} that the waiting times observed for earthquakes originating in Romania, Italy, United States of America (California region), and Japan~\cite{vivirschi2020}, as well as seismic events on the Moon~\cite{pgt2023}, exhibit a distinctive scale-free-like distribution. The aforementioned statistical results, as well as recent extensions on the distribution of motifs in earthquake networks~\cite{pana2023}, are supported by simple mechanical models, like the celebrated Olami-Feder-Christensen model~\cite{olami1992}, and suggest that seismic zones can be seen as self-organized critical systems. It should be noted that experimental data is not always supported by simple models and criticality is usually inferred through power-law distributions on some observables. In fact, there is a strong asymmetry between the large amount of statistical results obtained from the direct processing of empirical data and the substantially fewer results stemming from simplified structural models. This asymmetry reflects on one hand the large amounts of empirical data currently available, while on the other hand, it shows how difficult it is to construct simplified models of reduced computational load that can be used for large-scale statistical studies.

Motivated by our work on the available data pertaining to The International Centre for Advanced Studies on River-Sea Systems DANUBIUS-RI, a pan-European distributed research infrastructure supporting interdisciplinary research on River-Sea Systems~\cite{danubius2019}, we report here a similar analysis on one of the most extensive sea level catalogues. This dataset, covering more than a century of sea level data in the Port of Trieste, is an excellent example of Open Data and allows for a detailed investigation into the distribution of waiting times. In this article, a waiting time was defined as the time span between a given sea level $L$ and the next sea level of at least $L+\delta$ in the catalogue. Unlike earthquake databases, where we find seismic events recorded with varying degrees of precision on the position of the epicenters even for relatively recent events, the current catalogue offers very precise sea level data, with the imprecision on individual entries being less than 1 cm.

The dataset pertaining to the sea level in the Port of Trieste is one of the few ultra-centennial time series in the Mediterranean Sea~\cite{zerbini2017}. The evolution of the measurement mechanisms is explained in Ref.~\cite{rainich2023}. Briefly, the self-recording float tide gauge was first installed in 1859, when it was equipped with a stilling well opened in the floor of a room in the north-western corner of the Finance Guard building, at the end of Molo Sartorio, and was kept in operation until 1924. Then, after two years of renovations to the building, measurements were restarted in 1926 in a new tide-gauge hut, built on the same pier approximately 30 m to the east of the previous installation. Finally, in 1961, the hut was enlarged and a new stilling well was built, the measurements continuing without interruption until the present day.

The data used in our analyses is also described in Ref.~\cite{rainich2023} and openly available~\cite{seanoe2023} and covers the period 1905--2023. The quality of the recordings of the period 1859--1904 is questionable, as one can see from the inconsistencies between different reports, see Ref.~\cite{polli1947}  and Ref.~\cite{sterneck1905}. The data on the period 1905--1939 has been digitized from the original recordings, namely tabulations of hourly sea levels for 1905--1911 and 1913--1914 and charts from 1917 onward. The data from 1939 onward was already available and has only been revised in Ref.~\cite{rainich2023}. From 1905 to 2023, with the exception of the period between December 1924 and June 1926 when the tide gauge was not operational, there should be $1,034,377$ hourly recorded sea level values. Out of these, there are $3,531$ values, corresponding to $0.34\%$ of the entire dataset, which are estimated through interpolation of neighboring values, while $44764$, that is approximately $4.33\%$ of the dataset, are missing.

The rest of the article is structured as follows: in Section \ref{sec:method} we describe the methods used to determine the distribution of waiting times, while in Section \ref{sec:results} we present our numerical results both with and without data pre-processing. Lastly, Section \ref{sec:conclusion} gathers our concluding remarks and an outlook on future extensions.  

\section{Methods}
\label{sec:method}

The data catalogue detailed in the previous section was processed automatically, see Ref.~\cite{panagithub2023}, to determine a series of waiting times for given values of the sea level threshold $\delta$. For each chosen value of $\delta$ we sweep the dataset and at every value of sea level $L$ we determine the nearest subsequent data entry of value greater than or equal to $L+\delta$. The time span between these two sea levels is recorded in the waiting time series. Once the process is completed for all desired $\delta$ values, we determine the distributions of these numerically calculated waiting times, one distribution for each value of $\delta$, a process that is subject to some discussion in the specialized literature. As will be shown in Section \ref{sec:results}, these distributions have two distinct regimes, which are quite different: for small values of $\delta$ the distributions of waiting times have a very prominent scale-free character, while for the large values of $\delta$, the shape of the distribution is strongly impacted by data pre-processing. 
Please also note that these distributions reflect all the information available in the dataset, a situation which is quite different from similar computations on earthquake databases where small-magnitude events are discarded from the statistics.

Finally, let us add that our results stem solely from the statistical processing of recorded data~\cite{gao2023}, as we have not investigated the numerous (usually computationally demanding) models that can be used to describe -- either partly or in full -- the data, like those used for tides, storm surges, inverse barometric effect, river discharge, seasonal variability, and so on~\cite{muller2010, idier2019, muis2020}. Our approach is therefore data-driven with a focus on understanding to what extent one can predict certain events, particularly high sea level fluctuations. In this context, the distribution of waiting times is seen both as a statistical indicator of critical behavior (through the scale-free-like distribution of small values of $\delta$) and as a potential predictability indicator.

\subsection{Power-law distributed data}
\label{sec:fitting_power_law}

An interesting topic in the literature dedicated to fitting power-law distributed empirical data regards the distinction between directly fitting the probability density function, which can be altered by a variety of binning methods, and numerical methods that are better suited for the parameter estimation of a heavy-tailed distribution such as the maximum likelihood estimation (MLE) method ~\citep{clauset09}. The main problem is that binning methods can induce inaccurate estimates of the distribution parameters due to the (usually heavy) noise in the tail of the distribution (see, for instance, Ref.~\cite{Virkar2014}). Moreover, fitting data on logarithmic plots can lead to spurious errors in the value of the power-law exponent. Instead of directly fitting probability density functions subjected to binning procedures, an arguably better method is to represent the data via a cumulative distribution function (CDF) and compute the parameters by employing a goodness of fit estimator that operates on said CDF.

If one suspects that a given set of empirical data might be distributed as a probability function with a heavy tail, it is useful to plot the data using the complementary cumulative distribution function (CCDF), that is
\begin{equation}
    \hat{F}_X(x)=P(X>x)=1-F_X(x), 
\end{equation} where the right side of the equation shows the probability that the variable $X$ \textit{strictly exceeds} the value $x$.

Considering the probability distribution function (PDF) of a power law:
\begin{equation}
    p(x) = C x^{-\alpha},
\end{equation}
where $C$ is the normalization constant, the probability $P(x)$ that the variable has a value greater than $x$ is:
\begin{equation}
    P(x)=\int_x^\infty C(x')^{-\alpha}dx' = \frac{C}{\alpha-1}x^{-(\alpha-1)}. 
\end{equation}

The $\alpha$ exponent is computed by maximizing the log-likelihood function pertaining to the scale-free distribution with fixed parameter $x_{\mathrm{min}}$, which is the numerical implementation of the MLE method in the Python \texttt{powerlaw} package~\cite{alstott14}, with the added mention that further adjustment of the $\alpha$ exponent by the employed parameter optimization method in the near vicinity of the value supplied by the MLE method is possible.

The normalized expression for the power law is given by: 
\begin{equation}
    p(x) = \frac{\alpha-1}{x_{\mathrm{min}}}\left(\frac{x}{x_{\mathrm{min}}}\right)^{-\alpha}. 
\end{equation}
where $x_{\mathrm{min}}$ is the lower bound of the domain of values for $x$. The necessity of truncating the power-law at a value $x_{\mathrm{min}} > 0$ comes from the convergence condition imposed on the normalization integral which would diverge towards infinity otherwise.

\subsection{Goodness of fit}
\label{sec:goodness_of_fit}

For non-iid (independent and identically distributed) data, an appropriate measure for  quantifying the quality of fit is the Kolmogorov-Smirnov (KS) distance $D$~\cite{press1992}, as it determines the distance between two probability distributions, {\it i.e.},
\begin{equation}
    D = {\max}_{x \geq x_{\mathrm{min}}} |S(x)-P(x)|,
\end{equation}
which represents the maximum distance between the data and the fitted model CDFs. $S(x)$ denotes the CDF of the data for observations with values greater than or equal to $x_{\mathrm{min}}$ and $P(x)$ is the CDF corresponding to the power-law model that best fits the data considering $x \geq x_{\mathrm{min}}$. 
As is customary when comparing theoretical distributions with data obtained from real-world measurements, the CDF corresponding to the power law that has an input value range for $x$ spanning from $x_{\mathrm{min}}$ to $\infty$ will be renormalized using the appropriate area ratio taken with respect to the lower bound ($x_{\mathrm{min}}$) and upper bound (maximum observed waiting time) imposed by both the power-law and the observed data.

With the Kolmogorov-Smirnov distance serving as the measure for the goodness of fit, a multitude of parametric optimization methods can be employed in order to estimate the parameter pair ($\alpha,\, x_{\mathrm{min}}$) corresponding to a series of waiting times generated for a given $\delta$ value for which $D_{KS}$ is minimized. We mention, however, that due to the inherent sensitivity of the numerical procedure to the variations in the value of $x_{\mathrm{min}}$, parameter optimization methods that assume smoothness in the goodness of fit measure relative to the variation in the parameter values in order to employ a gradient or Hessian are not recommended, suitable alternatives being used instead ({\it e.g.}, Nelder-Mead, Simulated Annealing, Particle Swarm).
Furthermore, we note that any parametric optimization procedure will be heavily dependent on the initial values provided for ($\alpha,\, x_{\mathrm{min}}$). While the MLE method gives us suitable analytically computed values for $\alpha$ at any given $x_{\mathrm{min}}$, an initial $x_{\mathrm{min}}$ value that is too low will result in an unsatisfactory fit that avoids further adjustment of $x_{\mathrm{min}}$ and concentrates solely on estimating the best $\alpha$, while on the other hand an initial $x_{\mathrm{min}}$ that is too large will further seek to truncate an unnecessarily large portion of the data because it rapidly leads to a decrease in the Kolmogorov-Smirnov distance ({\it i.e.}, it is easier to improve the goodness of fit by having to fit less data than by trying to fit more data).

\subsection{Data pre-processing}
\label{sec:Data pre-processing}

In the current study, the pre-processing of data serves two main purposes: to take into account the cyclical behaviours embedded in the dataset by comparing the serialized data to a rolling average (a procedure called detrending) and to eliminate the noise generated by extreme values that do not contribute to the structure of the series of waiting times.

The first step employed in the pre-processing of raw data is the detrending. In order to achieve distributions of waiting times where trending behaviours are smoothed out, each data point of value $L$ from the raw dataset will have a corresponding value $L'$ in the pre-processed dataset that will be calculated as the relative difference between $L$ and the rolling average of $n$ data points preceding $L$, which will be expressed in percentage points (pps):
\begin{equation}
    L' = \frac{L - RA}{RA} \cdot 100
\end{equation}
where $RA$ denotes the rolling average that is computed at each corresponding $L$ value. In order to avoid accidental division by 0 (encountered when both positive and negative contributions  to the rolling average cancel each other out), the whole raw dataset is shifted prior to the computation of $L'$ values such that no value of $L$ is less than 1. This intermediary operation is justified because the final values of $L'$ are ultimately expressed as percentage points fluctuations from a rolling average computed from the already shifted data.

In the particular case of sea level variations, a time interval of 29.5 days was considered as the period of time that covers the span of values used in the computation of the rolling average, which converted in hours gives us a number of $n=708$ data points. The specific duration of 29.5 days was chosen because it corresponds to the average period characterizing lunar phases to which tidal movements are heavily influenced. 
It is important to note that this numerical procedure, while smoothing out certain oscillatory components locally embedded in the data, succeeds in conserving the global structure of the waiting time distributions that constitute the main focus of any further analysis. This is achieved because the technique does not make use of any additional information from outside of the dataset, and it does not infer any knowledge about the local variation of the data values ({\it e.g.}, as compared to interpolation procedures).

The second part of the data pre-processing is the elimination (or pruning) of extreme values. In the current study, the top 0.04\% of extreme values (both positive and negative fluctuations) were pruned from the pre-processed dataset from which the distributions of waiting times are derived. What this achieves, is the elimination of noise in the region of extremely large waiting times that deviate significantly from the structure of the distribution's heavy tail. The elimination of such noise from the analyzed distributions not only increases the accuracy of the fit (by eliminating its contribution towards the normalization of the distributions) but also decreases the computational cost incurred by generating the corresponding waiting times associated with it.

\section{Results}
\label{sec:results}

\subsection{Distributions of waiting times pertaining to the raw dataset}

We first focus on the results of the distribution of waiting times using the raw data, without pre-processing, and show that for small values of $\delta$ the distribution is akin to a scale-free one, while for large values of $\delta$ it resembles an exponential distribution. 

\begin{figure}[hbt!]
    \centering
    \begin{subfigure}[b]{0.49\textwidth}
         \centering
         \includegraphics[width=\textwidth]{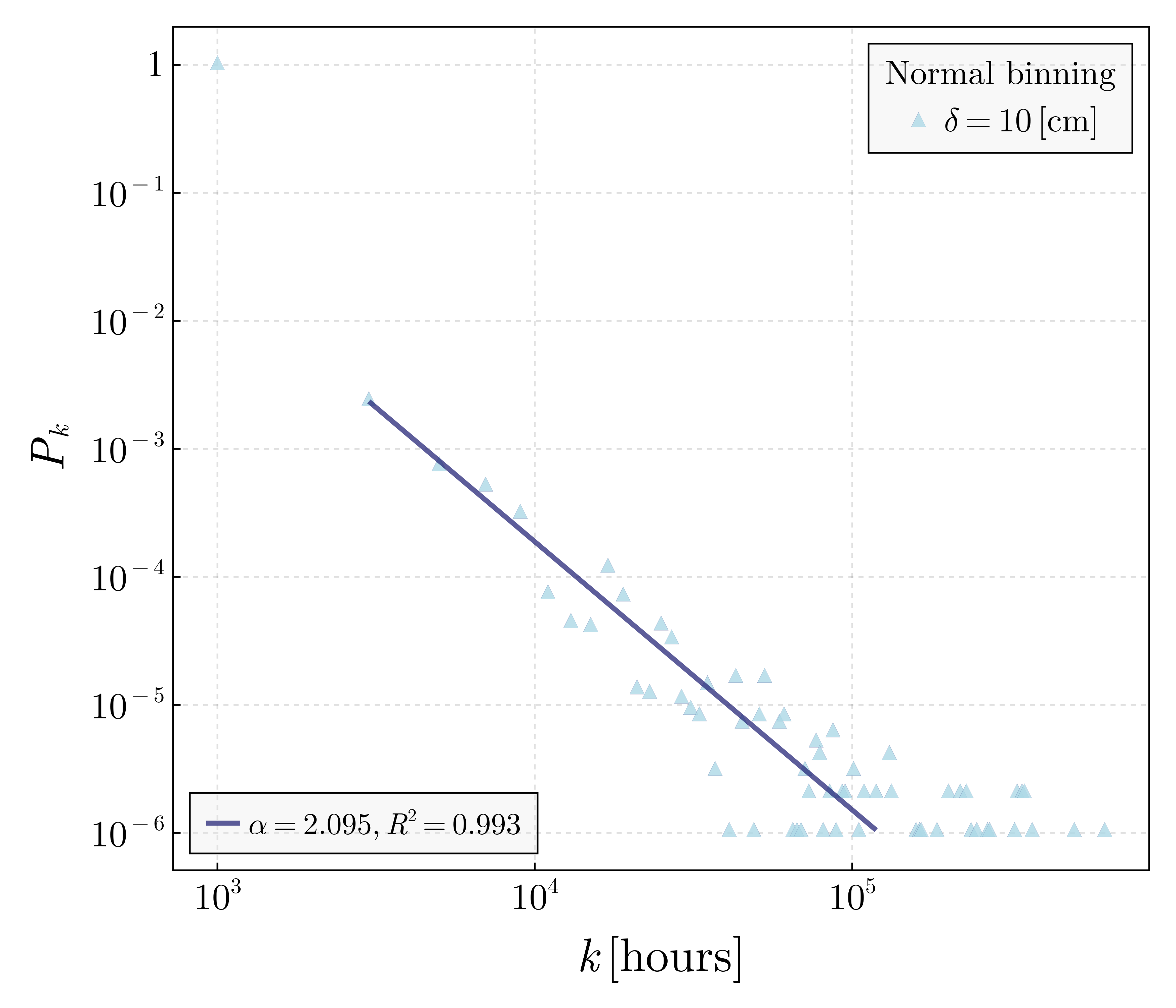}
         \caption{Linear binning of PDF}
         \label{fig:normbin}
    \end{subfigure}
    \hfill
    \centering
    \begin{subfigure}[b]{0.49\textwidth}
         \centering
         \includegraphics[width=\textwidth]{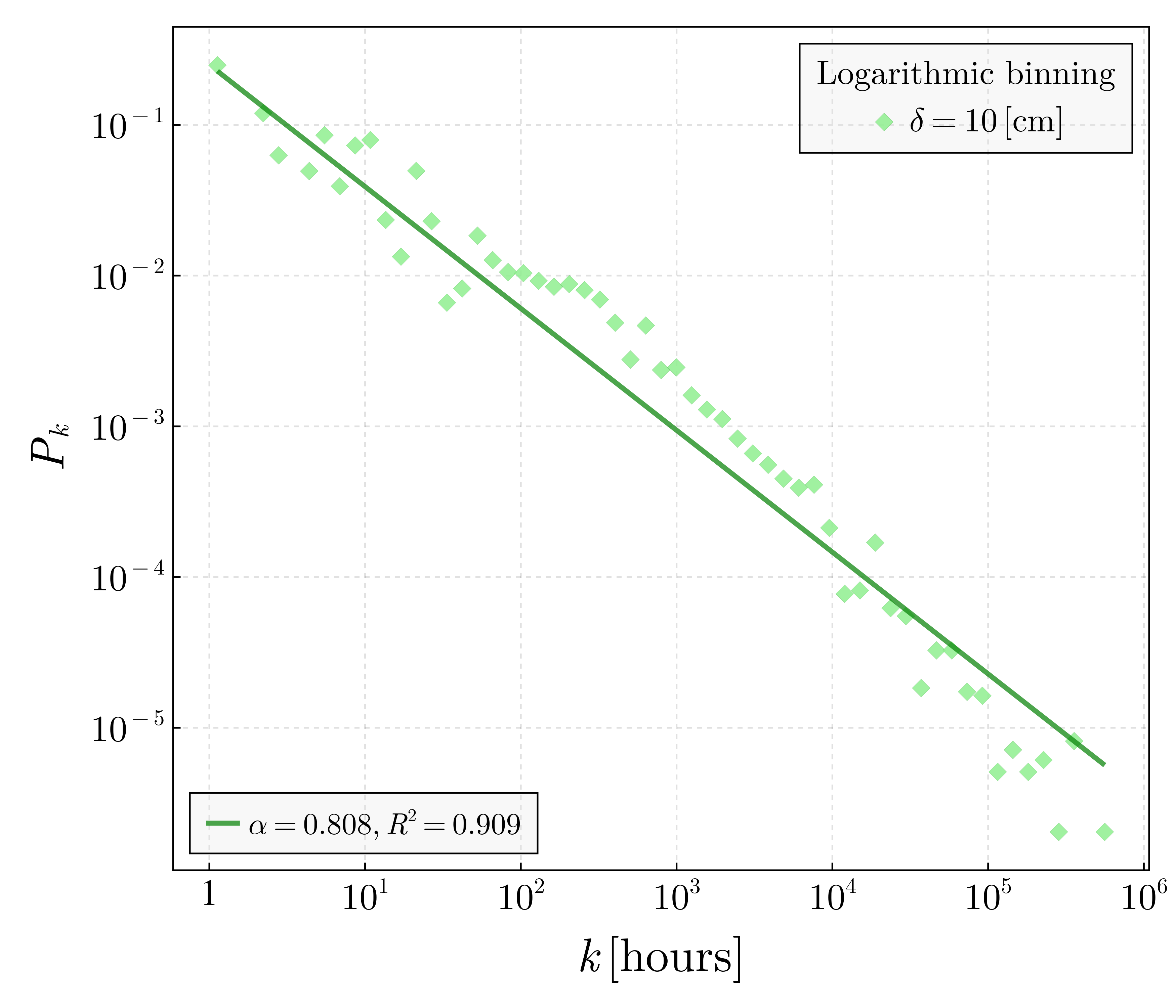}
         \caption{Logarithmic binning of PDF}
         \label{fig:logbin}
    \end{subfigure}
    \\
    \centering
    \begin{subfigure}[b]{0.49\textwidth}
         \centering
         \includegraphics[width=\textwidth]{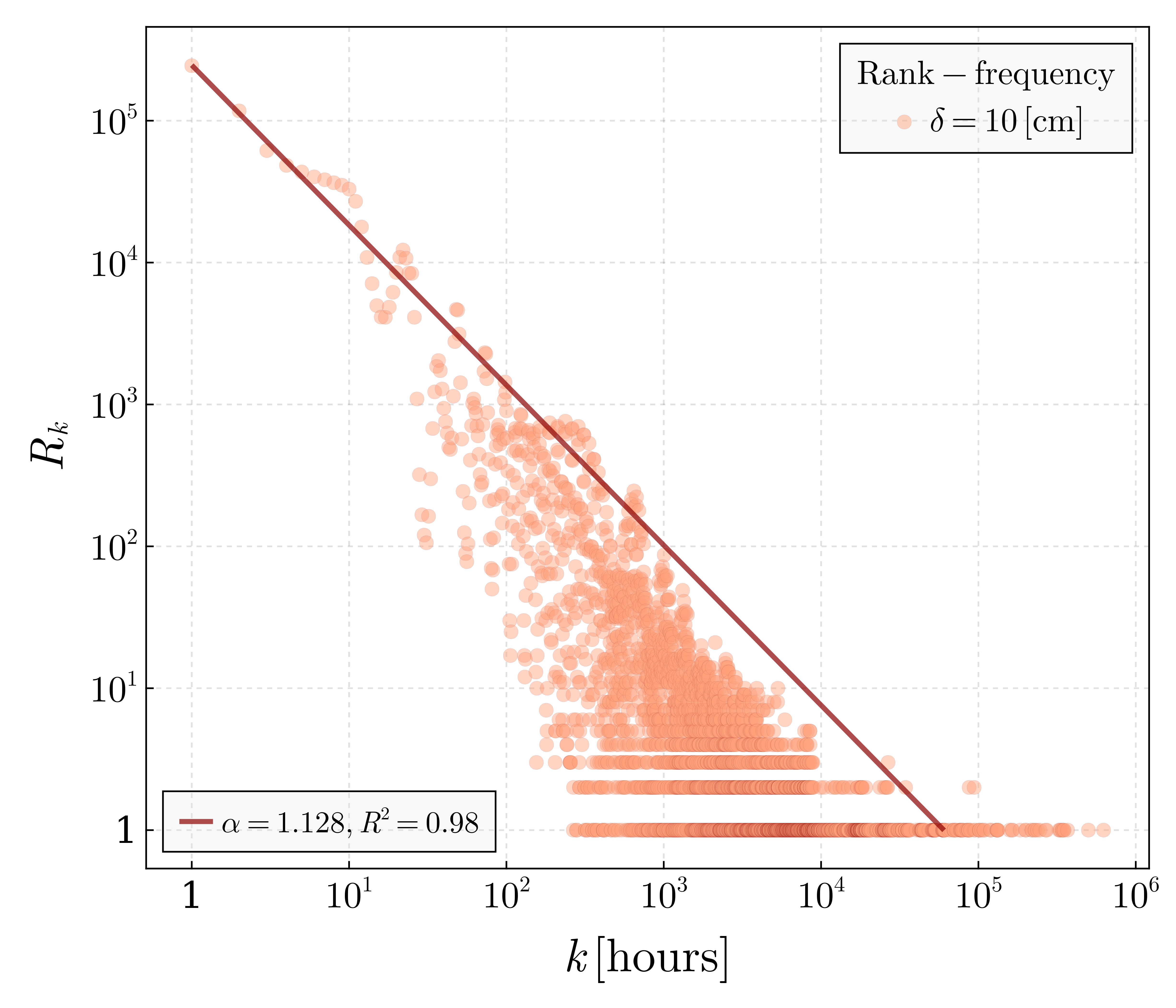}
         \caption{Rank-frequency plot}
         \label{fig:rf}
    \end{subfigure}
    \hfill
    \centering
    \begin{subfigure}[b]{0.49\textwidth}
         \centering
         \includegraphics[width=\textwidth]{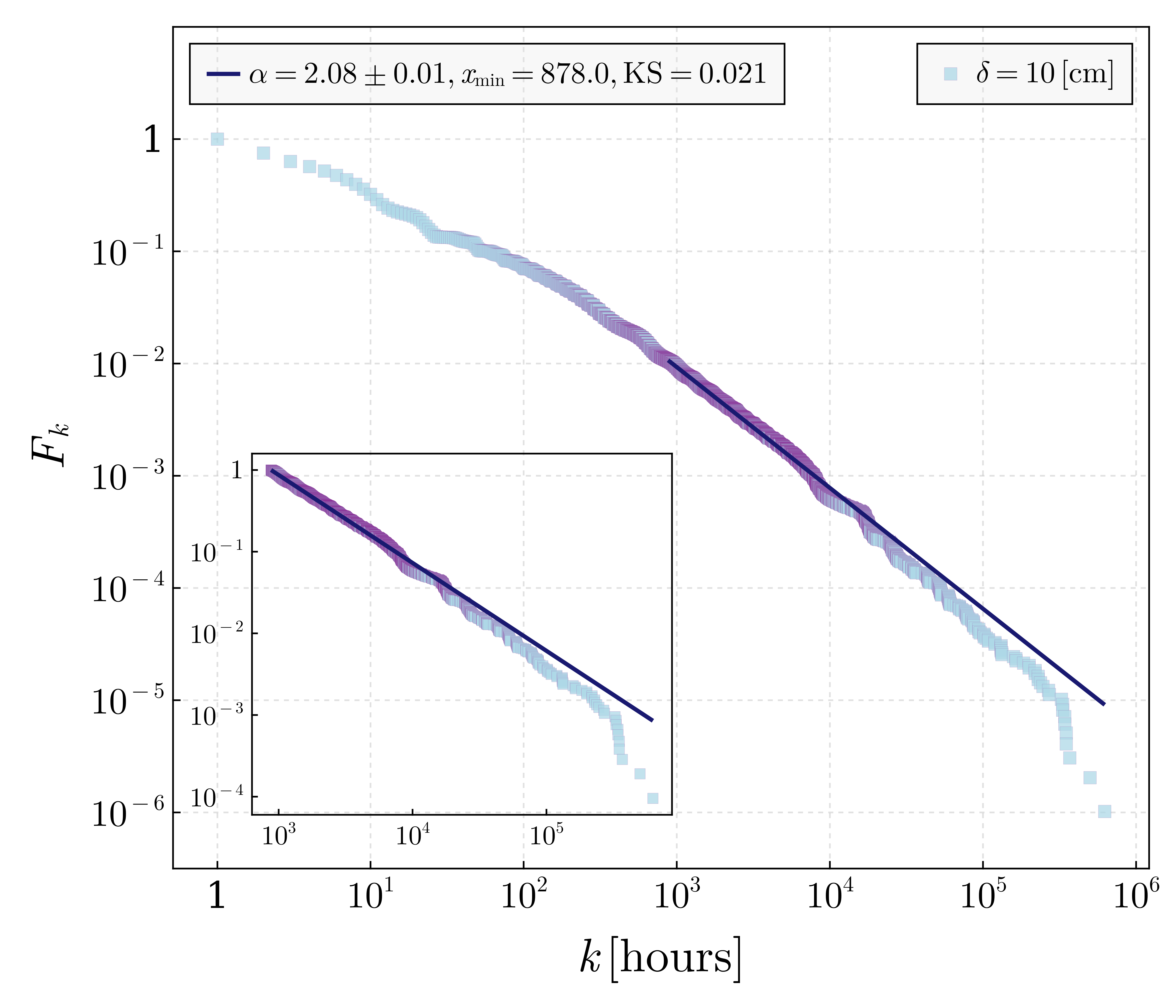}
         \caption{CCDF representation}
         \label{fig:mle}
    \end{subfigure}

    \caption{Distribution of waiting times for $\delta=10$ cm obtained by: fitting the PDF with linear binning (panel (a)), fitting the PDF with logarithmic binning (panel (b)), fitting the rank-frequency distribution (panel (c)) and employing the Kolmogorov-Smirnov parameter optimization (panel (d)). Please note that while the observed data exhibits a clear scale-free behaviour in all panels, there are substantial differences in the fitted value of the $\alpha$ exponent due to the fact that directly fitting the probability density function (with least squares in this case) yields results that accommodate the data at low waiting times at the detriment of data situated at the heavy tail of the distribution. This numerical sensitivity of the fitting results to the binning of the probability distribution is successfully bypassed by the parametric optimization employing the Kolmogorov-Smirnov distance. In all panels, the results pertain to the entire 1905--2023 dataset.}
    \label{fig:bin_vs_mle}
\end{figure}

\begin{figure}[]
    \centering
     \includegraphics[width=.55\textwidth]{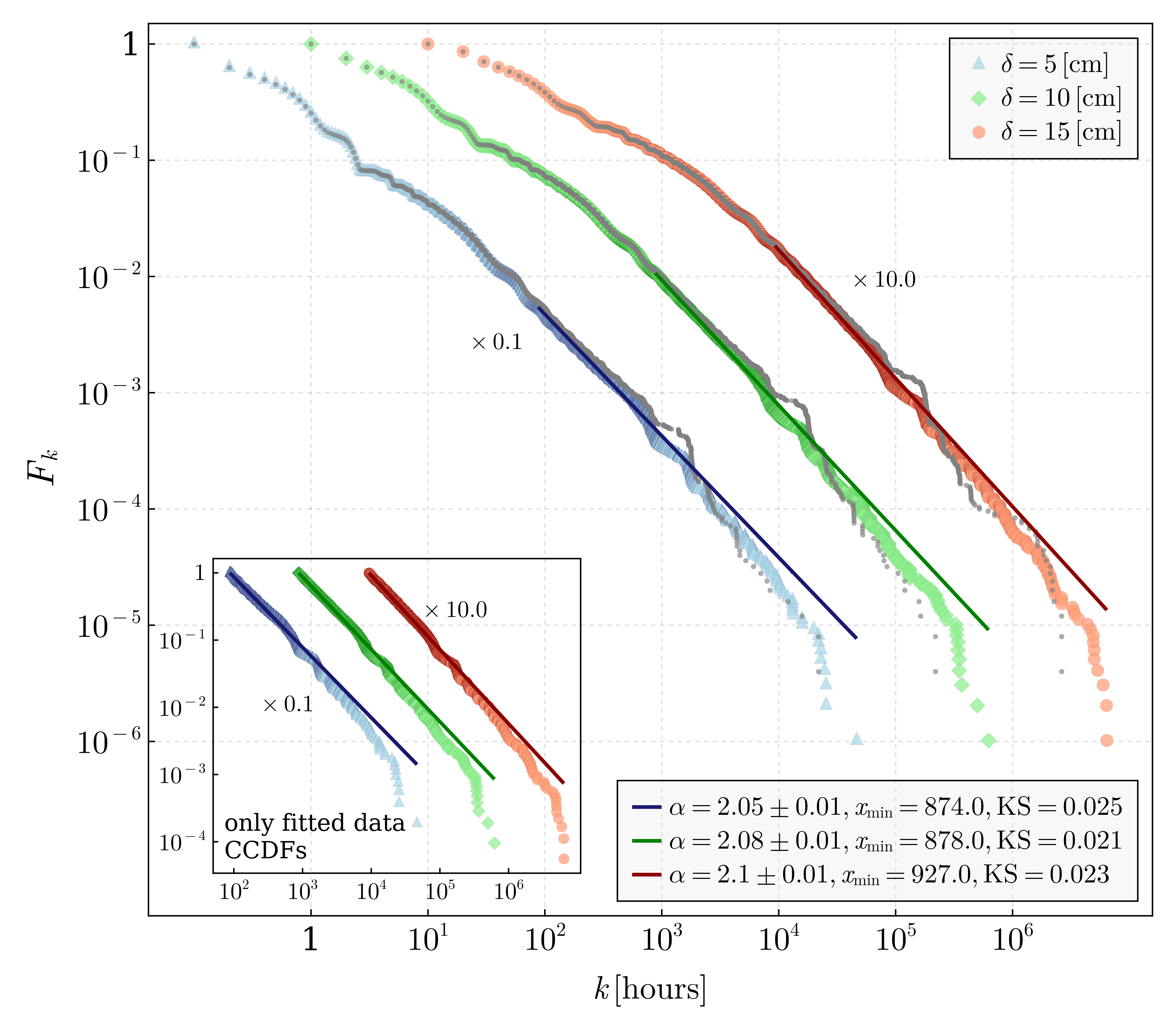}
     \caption{Distributions of waiting times for $\delta$ = 5, 10, and 15 cm. The three plots have been horizontally shifted for extra clarity. Please note that the three distributions are qualitatively identical, the only quantitative difference being with the $\alpha$ exponent, which increases with $\delta$. Please note that the colour curves correspond to the entire 1905--2023 dataset, for which the $\alpha$ exponent was also computed, while the black dotted lines show the distribution obtained using the 1939--2023 dataset.}
     \label{fig:best_fits}
\end{figure}

\begin{figure}[hbt!]
    \centering
    \begin{subfigure}[b]{0.49\textwidth}
         \centering
         \includegraphics[width=\textwidth]{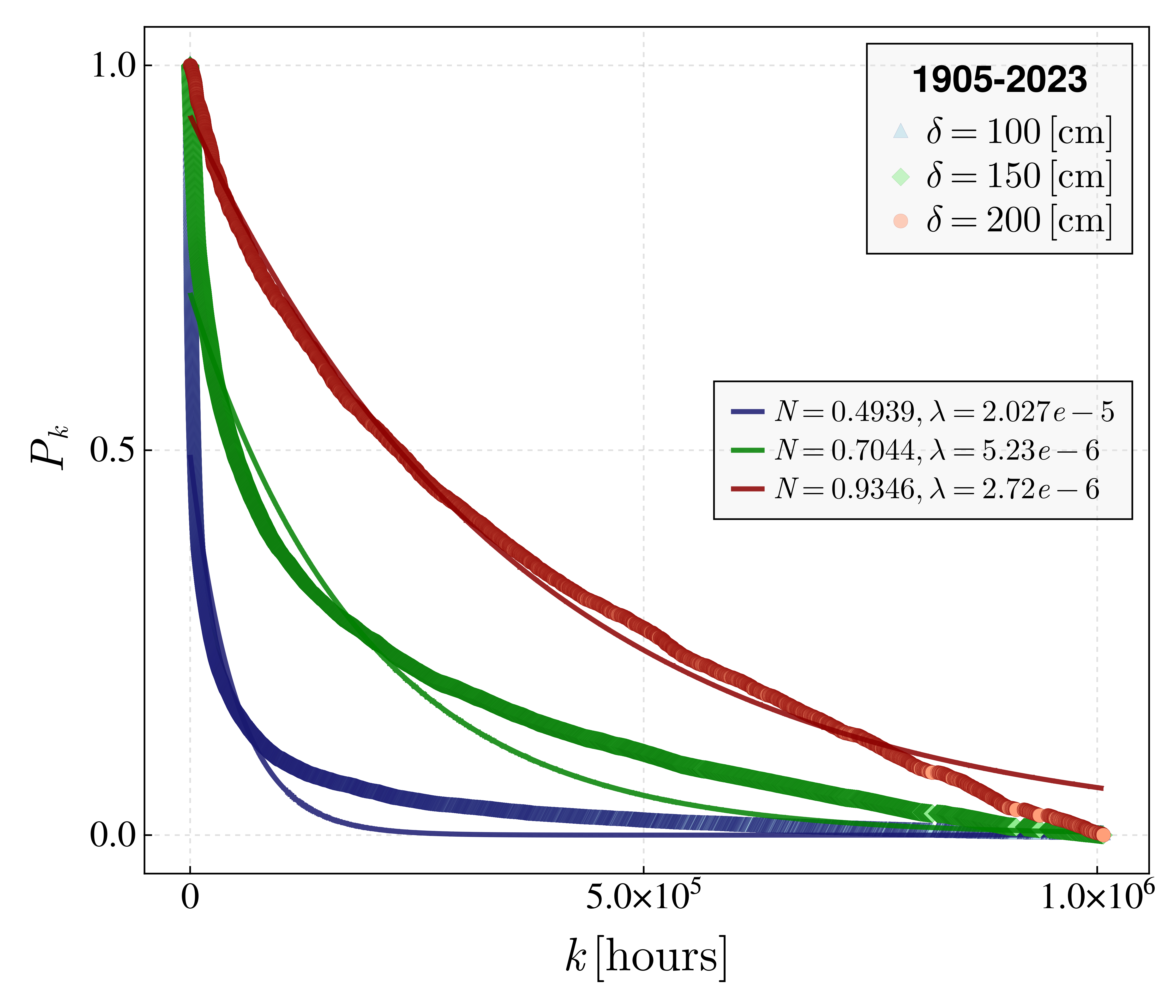}
         \caption{1905-2023}
         \label{fig:large_delta_exp_all_data}
    \end{subfigure}
    \hfill
    \centering
    \begin{subfigure}[b]{0.49\textwidth}
         \centering
         \includegraphics[width=\textwidth]{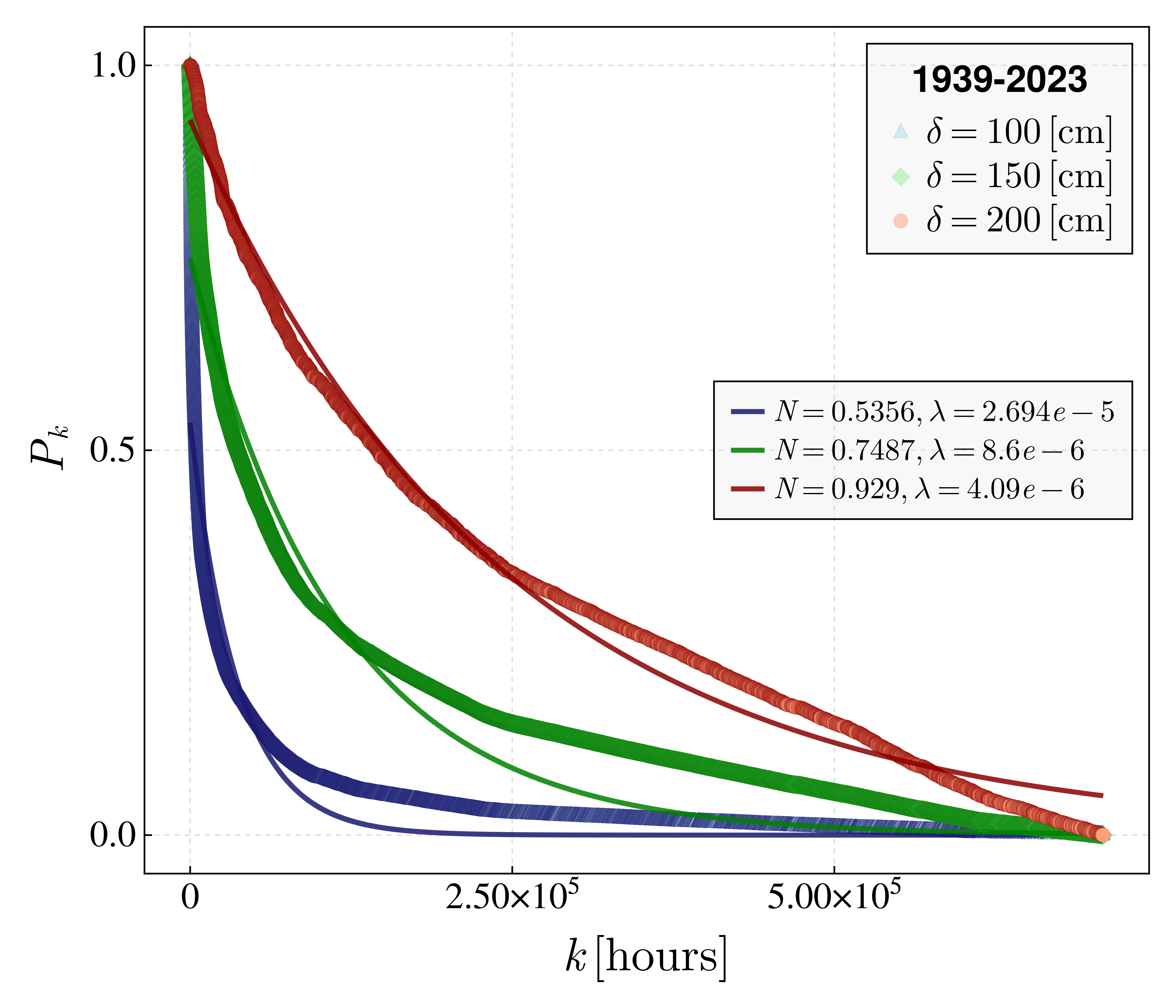}
         \caption{1939-2023}
         \label{fig:large_delta_exp_non_historic}
    \end{subfigure}

    \caption{Distribution of waiting times in the limit of large values of $\delta$. The left panel corresponds to the entire 1905--2023 dataset, while the right panel shows the results using only the 1939--2023 dataset. The distributions observed in both panels are exponential-like, the only differences being in the fine structure of the distribution at long waiting times.}
    \label{fig:large_delta_exp}
\end{figure}

Thus, we depict in Fig. \ref{fig:bin_vs_mle} a typical distribution of waiting times to illustrate the differences between results obtained through typical fitting procedures on binned probability density functions (panels (a), and (b)), a rank-frequency plot (panel (c)) and the numerical recipe employing the Kolmogorov-Smirnov distance (panel (d)). The main point is that binning methods introduce a rather spurious numerical noise, especially at long waiting times, which, in turn, makes the fitting of the $\alpha$ exponent rather imprecise, to the extent that for logarithmic binning we observe an unphysical $\alpha$ scaling exponent smaller than 1. Moreover, different binning strategies generate different fits with varying levels of accuracy, as measured, for instance, through $R^2$ (see the info in panels (a), (b), and (c) of Fig. \ref{fig:bin_vs_mle}). While the reported results were obtained using the entire dataset, virtually identical results are obtained when considering either the 1905--1939 subset or the 1939--2023 one.

In Fig. \ref{fig:best_fits} we present different waiting times distributions, obtained for low values of $\delta$, namely 5, 10, and 15 cm, to show that these distributions are indeed scale-free-like over a few orders of magnitude and that this result is not accidental. Please note that the $\alpha$ exponents have very similar values and that the three plots have been intentionally spread for clarity. We would like to stress that these results are very robust with respect to errors in the dataset well above the maximal sea level imprecision of 1 cm. We have checked, for instance, the changes in the $\alpha$ exponent induced by artificially generated sea level errors of up to 5\% of the maximal sea level and noticed that the distribution of waiting times retains its scale-free character, the only change being in the $\alpha$ exponent itself which varies slightly.

Lastly, we show in Fig. \ref{fig:large_delta_exp} the results for large values of $\delta$, ranging from 100 to 200 cm. As seen from the figure, the distributions are no longer scale-free-like and become, in fact, exponential-like distributions. Let us also add that, unlike the small $\delta$ results, the waiting times in Fig. \ref{fig:large_delta_exp} are slightly impacted by the inclusion or exclusion of the historical dataset, as can be easily observed by comparing the left panel (which pertains to the entire 1905--2023 period) with the right one (which pertains to the 1939--2023 period). The overall observed distributions are still exponential-like but a close inspection of the depicted curves shows that the fine structure exhibits some differences, particularly for long waiting times.

\subsection{Distributions of waiting times for the pre-processed data}

The numerical procedure described in \ref{sec:Data pre-processing} was used on the raw dataset values spanning from 1927 to 2023 because 1926 was the last year where significant gaps in the dataset exist. We consider an 708--hour rolling average, corresponding to a 29.5-day window, to smooth out the hierarchy of small time-scales up to the 29.5 days, which corresponds to the average period characterizing lunar phases.

\begin{figure}[]
    \centering
    \begin{subfigure}[]{0.49\textwidth}
         \centering
         \includegraphics[width=\textwidth]{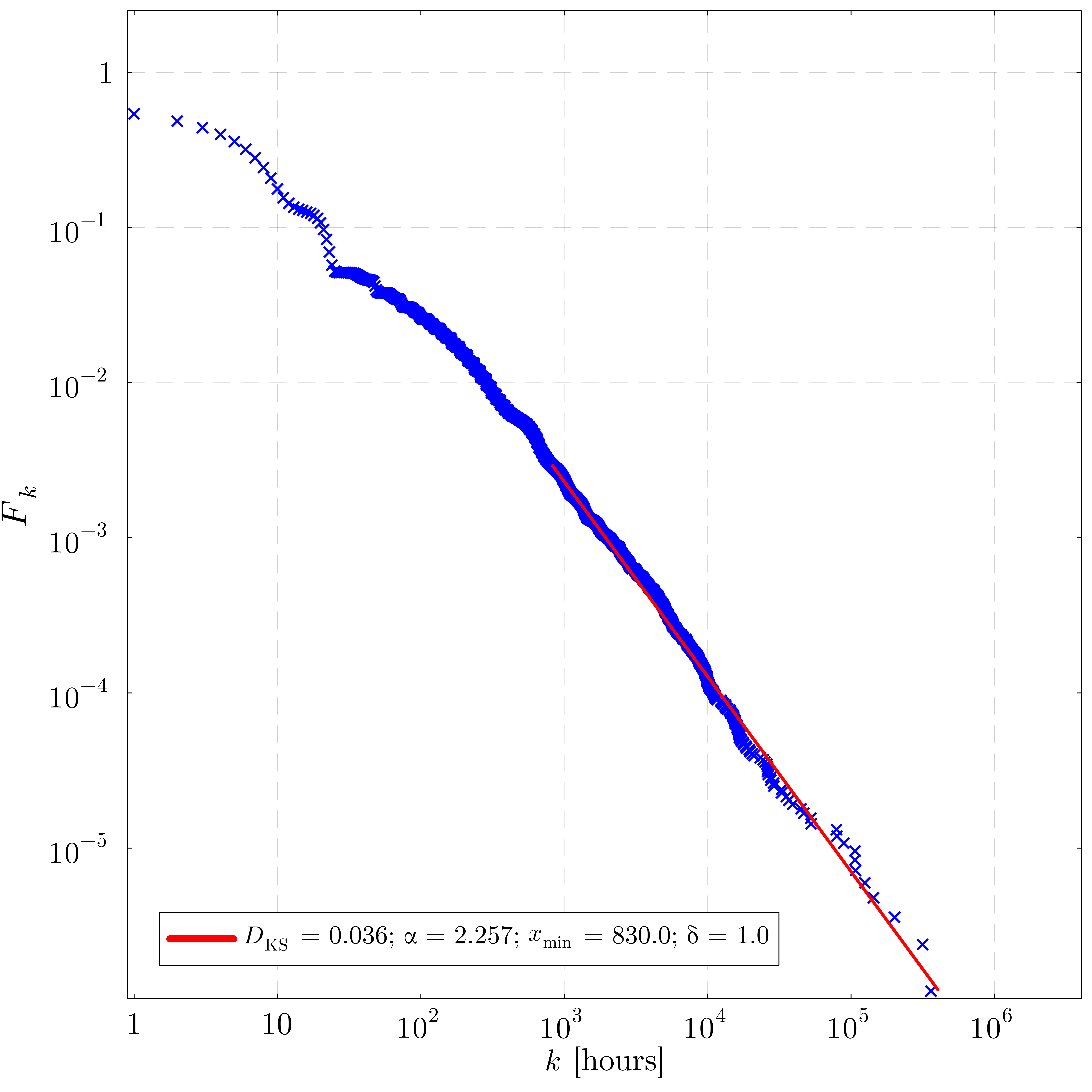}
    \end{subfigure}
    \hfill
    \centering
    \begin{subfigure}[]{0.49\textwidth}
         \centering
         \includegraphics[width=\textwidth]{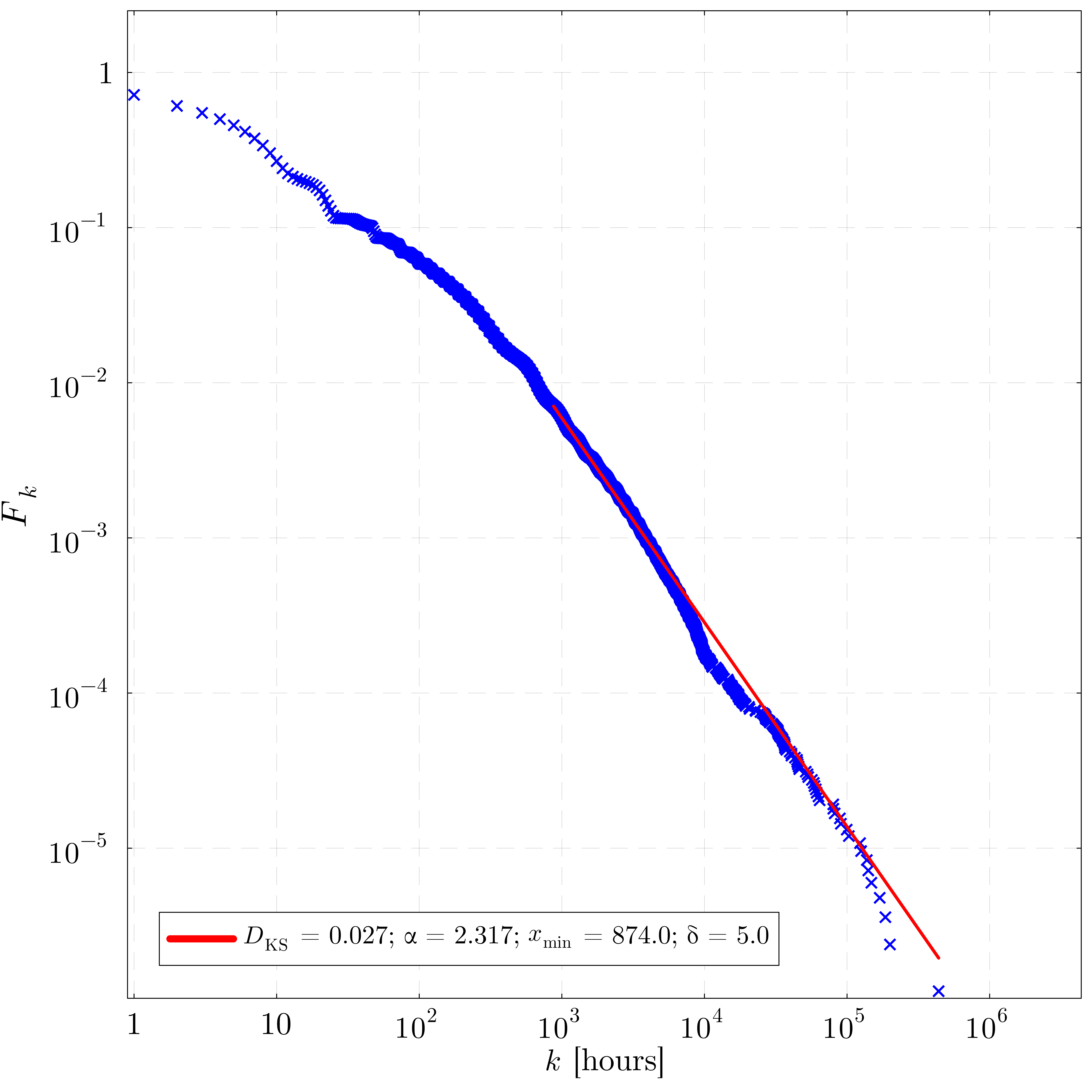}
    \end{subfigure}
    \\
    \centering
    \begin{subfigure}[]{0.49\textwidth}
         \centering
         \includegraphics[width=\textwidth]{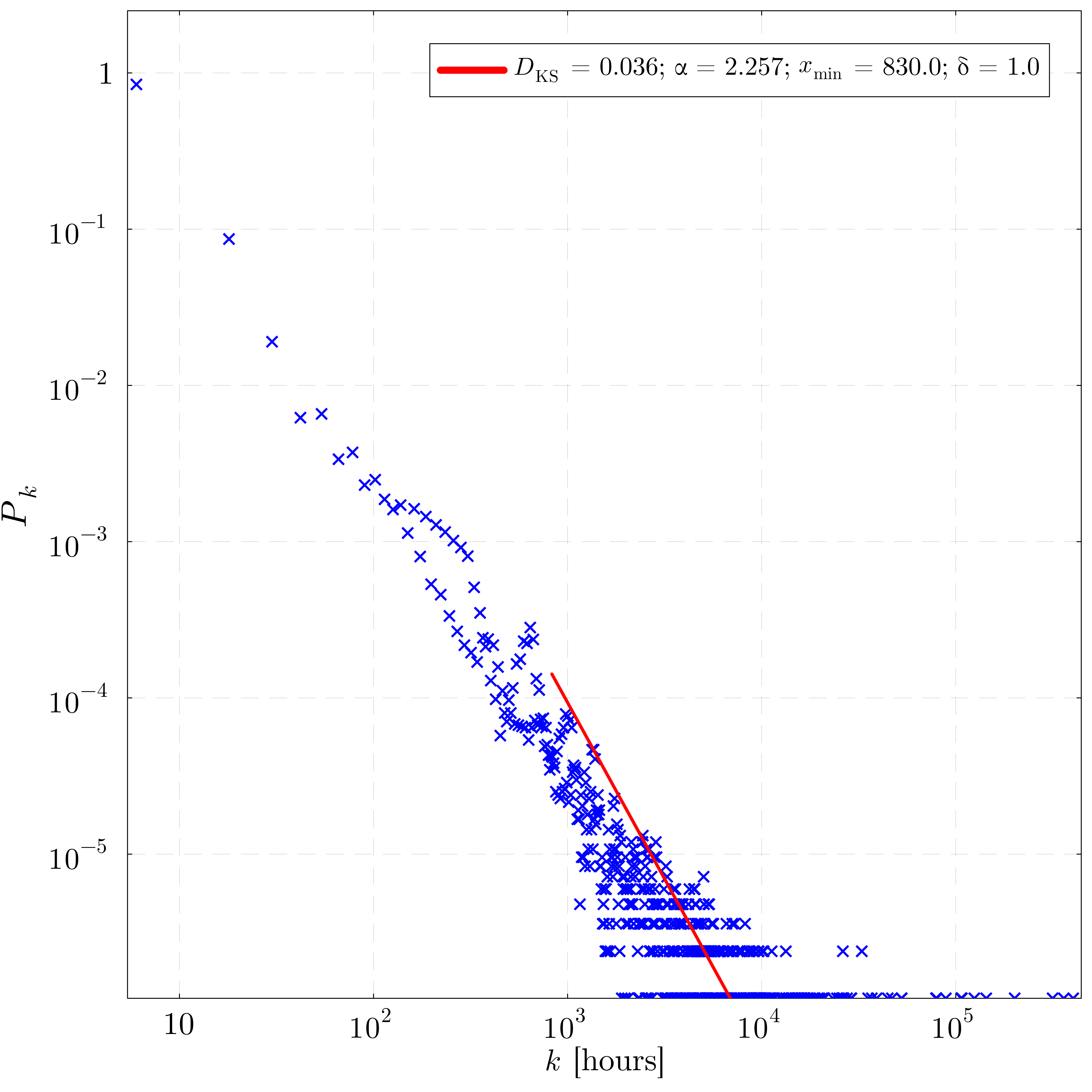}
    \end{subfigure}
    \hfill
    \centering
    \begin{subfigure}[]{0.49\textwidth}
         \centering
         \includegraphics[width=\textwidth]{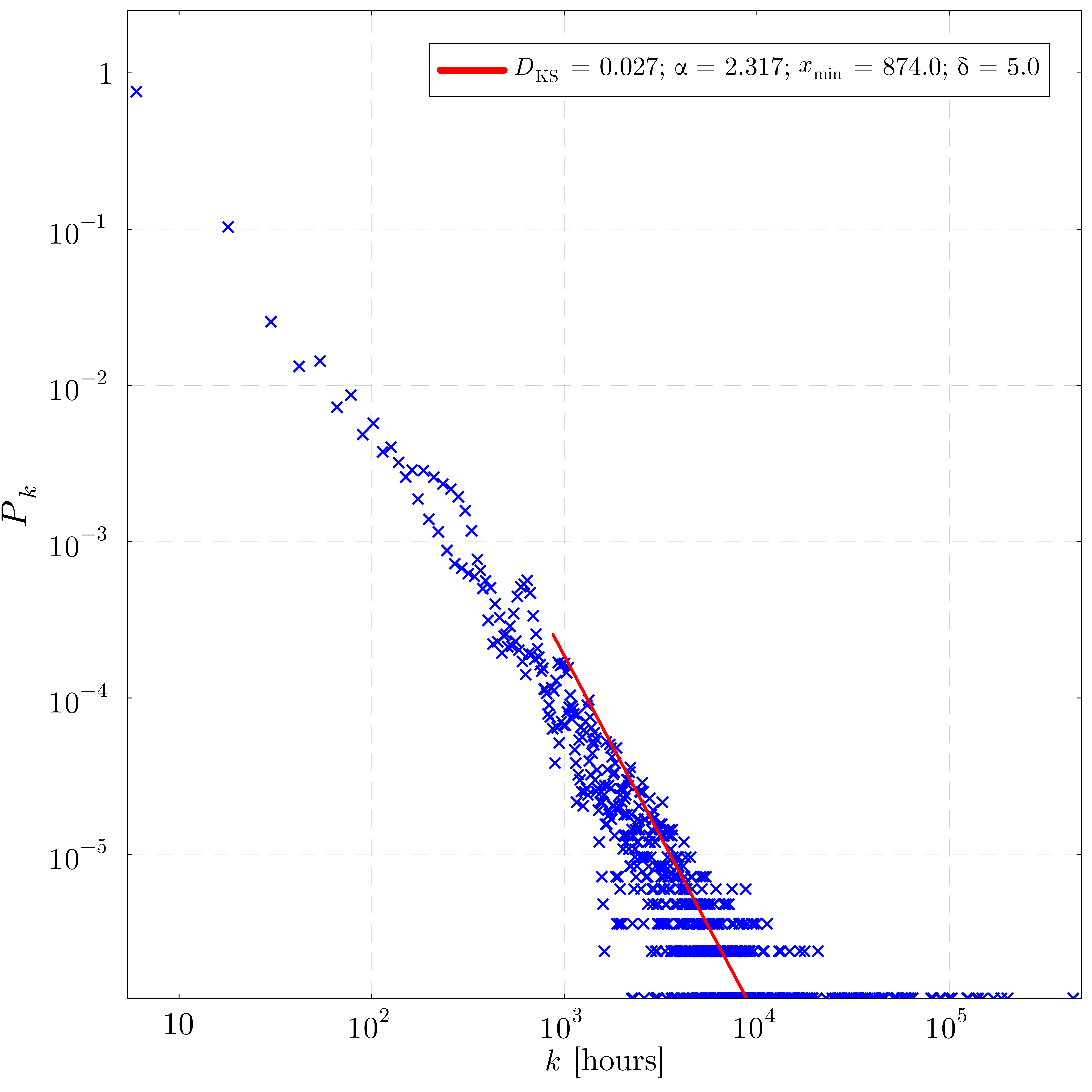}
    \end{subfigure}

    \caption{Distribution of waiting times for $\delta$ = 1 pps (left side figures) and  $\delta$ = 5 pps (right side figures). Figures on the top row show the representation of the CCDF while on the bottom row, the corresponding PDF plots are shown with a 12-hour linear binning applied. The parametric optimization fits were obtained by employing the Kolmogorov-Smirnov numerical recipe discussed in Section \ref{sec:goodness_of_fit}.}
    \label{fig:pre-processed SF plots}
\end{figure}

\begin{figure}[]
    \centering
    \begin{subfigure}[]{0.49\textwidth}
         \centering
         \includegraphics[width=\textwidth]{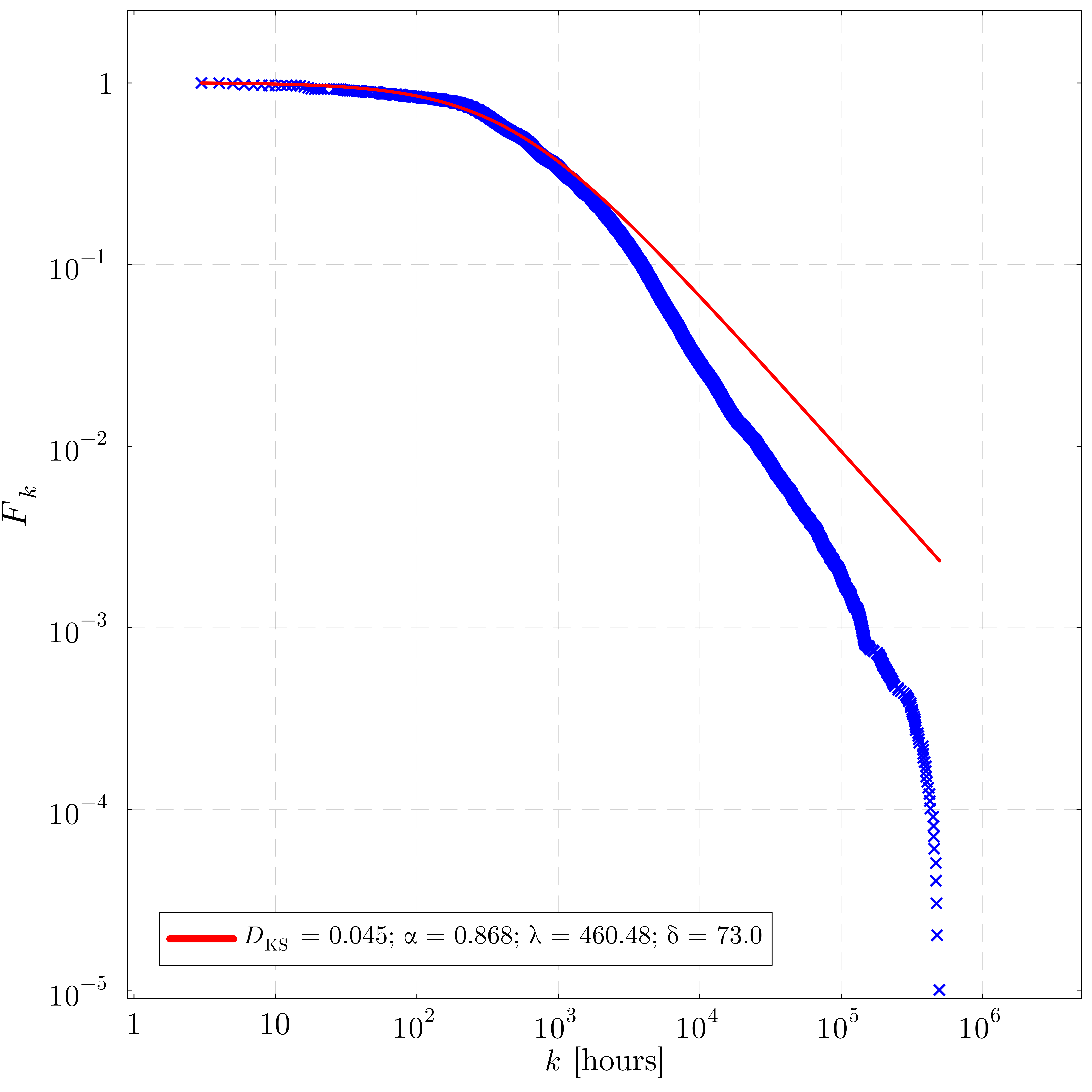}
    \end{subfigure}
    \hfill
    \centering
    \begin{subfigure}[]{0.49\textwidth}
         \centering
         \includegraphics[width=\textwidth]{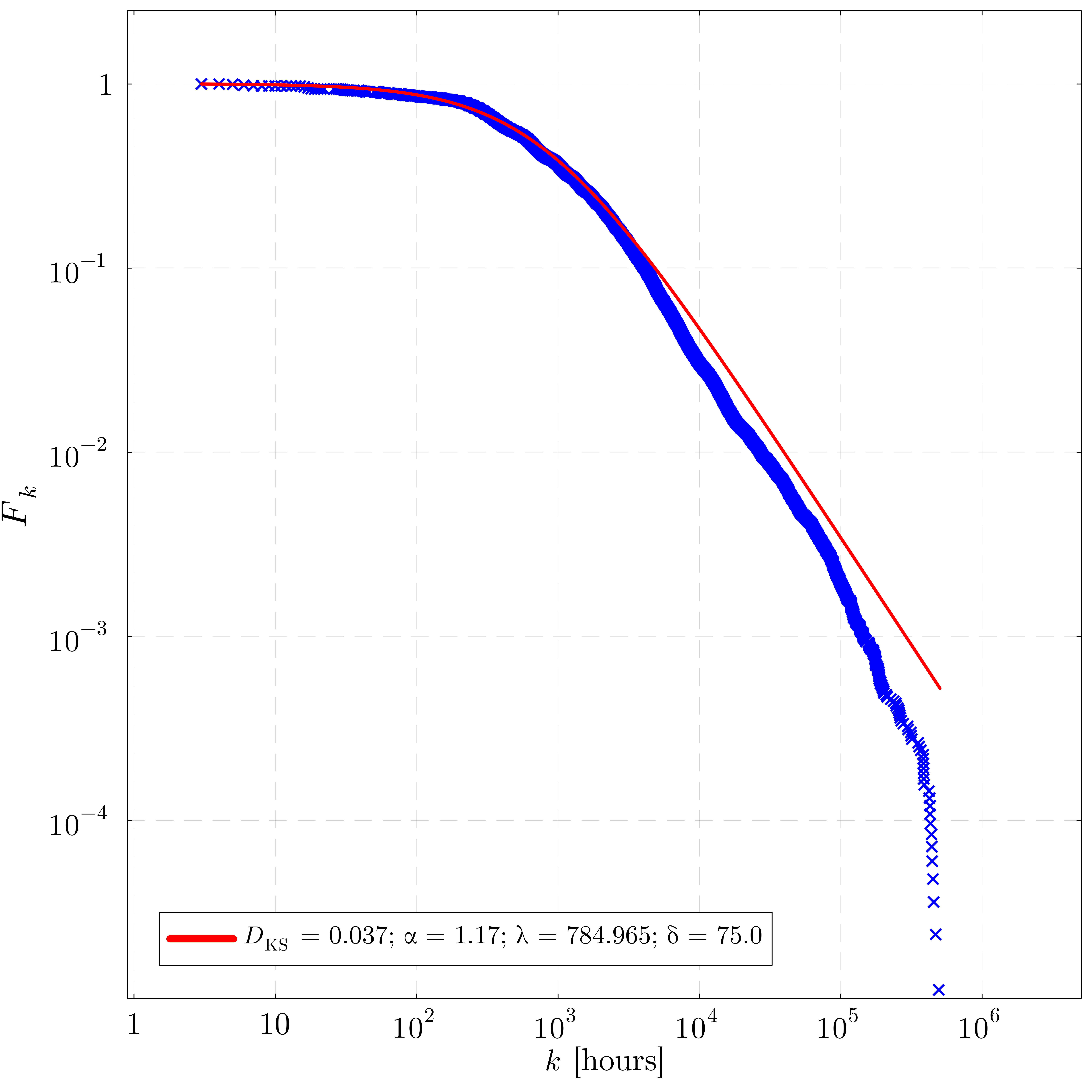}
    \end{subfigure}
    \\
    \centering
    \begin{subfigure}[]{0.49\textwidth}
         \centering
         \includegraphics[width=\textwidth]{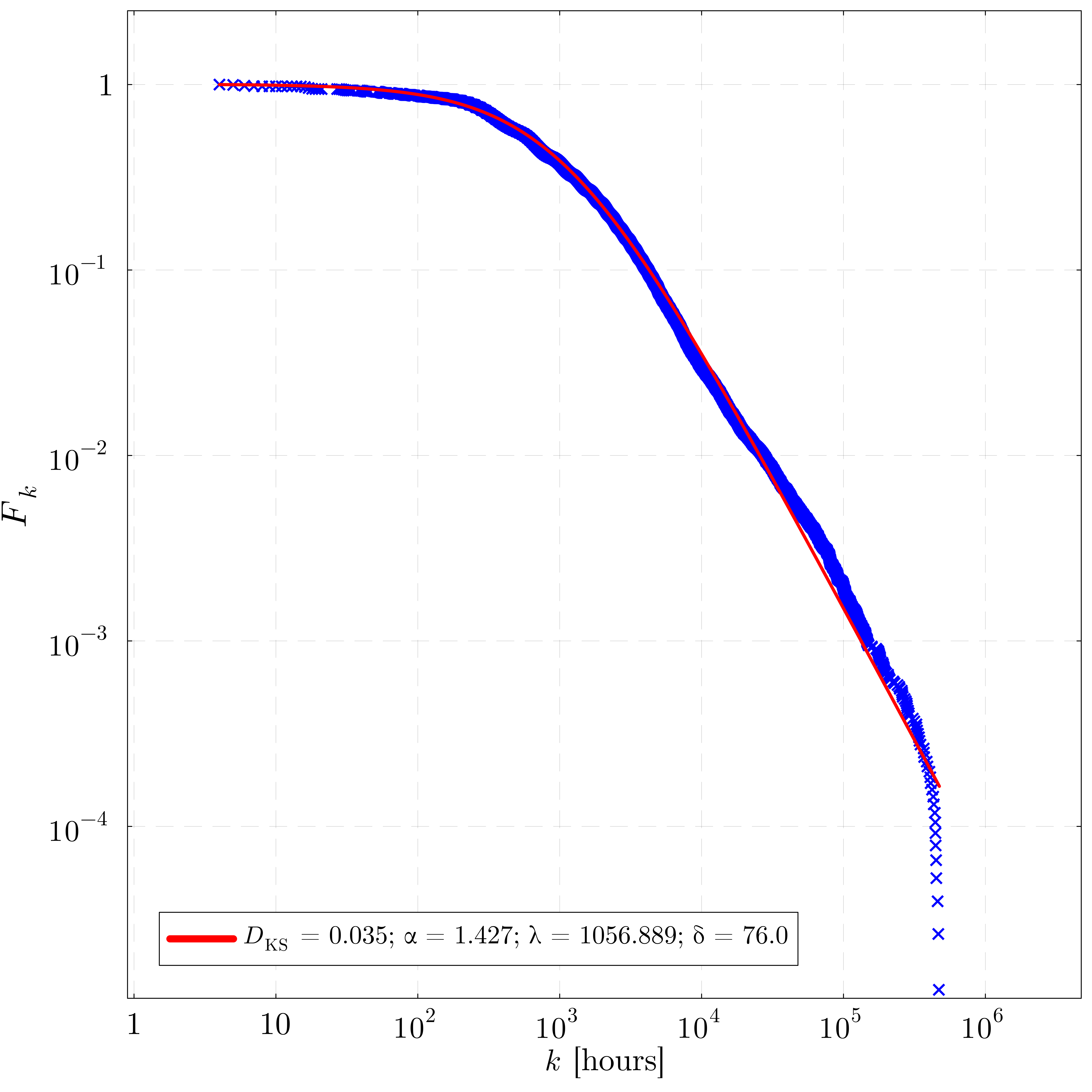}
    \end{subfigure}
    \hfill
    \centering
    \begin{subfigure}[]{0.49\textwidth}
         \centering
         \includegraphics[width=\textwidth]{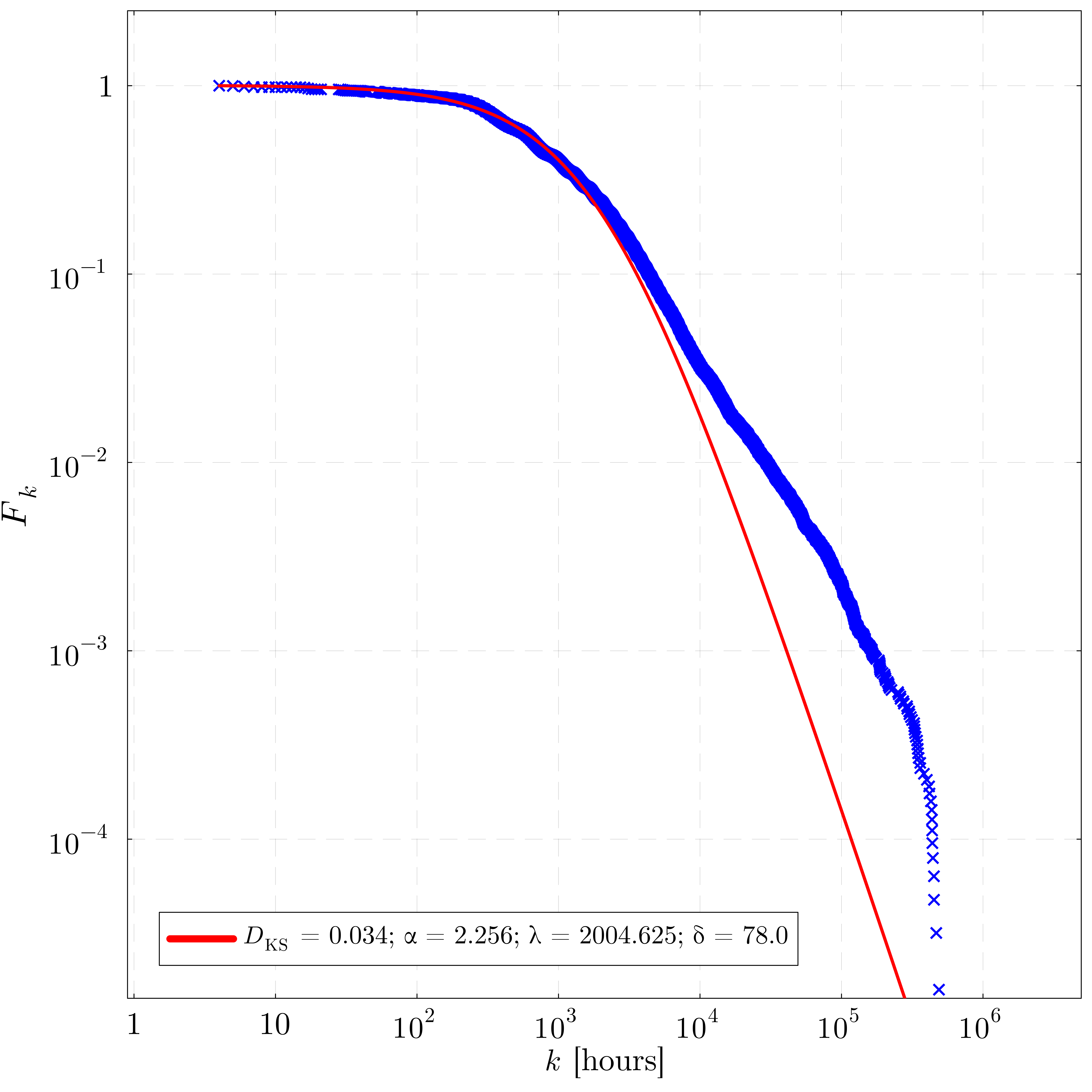}
    \end{subfigure}

    \caption{Distribution of waiting times for $\delta$ = 73, 75, 76 and 78 pps fitted with Pareto-Tsallis type functions. The CCDF representations clearly show that the fitted distributions at $\delta$ = 73 and 75 pps overshoot the data in the region of the heavy tail, the distribution represented at $\delta$ = 78 pps undershoots the data in the same region, while the fitted distribution pertaining to $\delta$ = 76 pps describes the data accordingly. The parametric optimization fits were obtained by employing the Kolmogorov-Smirnov numerical recipe discussed in Section \ref{sec:goodness_of_fit}.}
    \label{fig:pre-processed PT CCDF plots}
\end{figure}

Bearing in mind that the quantitative meaning of $\delta$ has shifted from observed sea level measurements expressed in cm to relative fluctuations of those measurements from a 708--hour rolling average of past data values (expressed in percentage points or pps), we examine the behaviour of the waiting time distributions at both small and large $\delta$ values. In Fig. \ref{fig:pre-processed SF plots} we show that the waiting time distributions retain their scale-free behaviour for small $\delta$ values after pre-processing is applied, thus showing that the global structure of the distribution is maintained.

\begin{figure}[]
    \centering
    \begin{subfigure}[b]{0.49\textwidth}
         \centering
         \includegraphics[width=\textwidth]{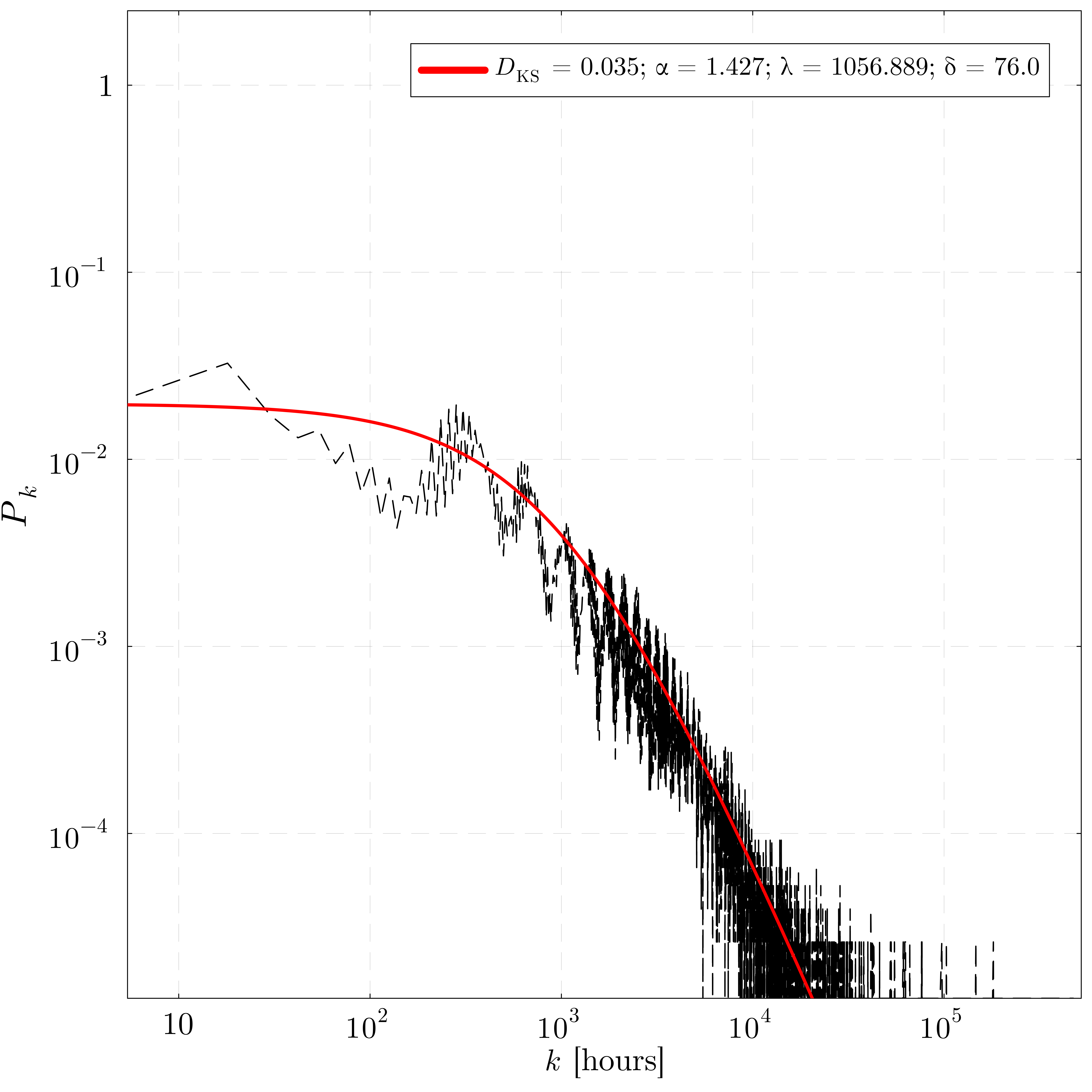}
         \caption{PDF: 12-hour linear binning}
    \end{subfigure}
    \hfill
    \centering
    \begin{subfigure}[b]{0.49\textwidth}
         \centering
         \includegraphics[width=\textwidth]{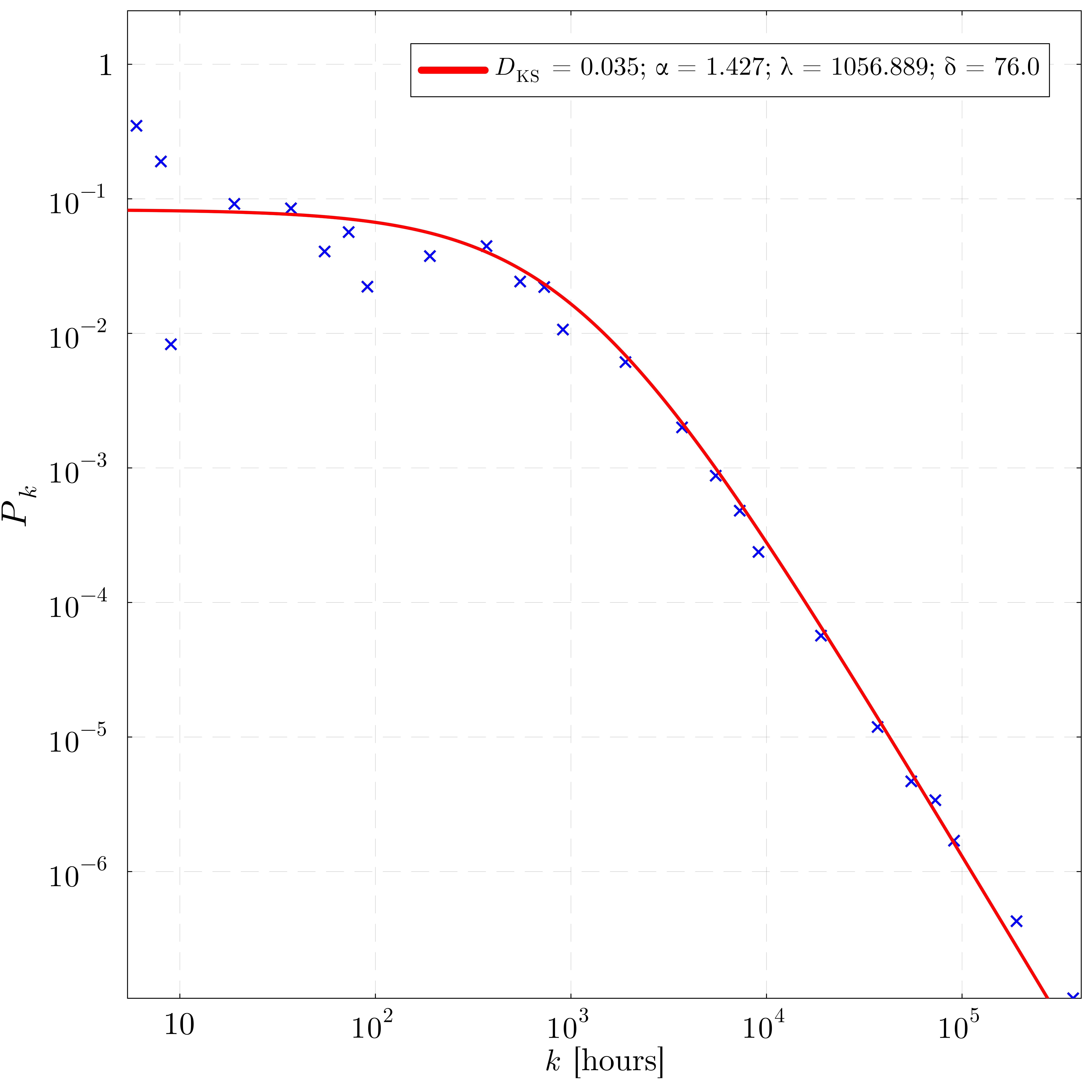}
         \caption{PDF: logarithmic binning}
    \end{subfigure}

    \caption{Distribution of waiting times for $\delta$ = 76 pps fitted with Pareto-Tsallis type functions. The binned PDF representations are those corresponding to the CCDF plot in Fig. \ref{fig:pre-processed PT CCDF plots}. The parametric optimization fits were obtained by employing the Kolmogorov-Smirnov numerical recipe discussed in Section \ref{sec:goodness_of_fit}.}
    \label{fig:pre-processed PT PDF plots}
\end{figure}

In Fig. \ref{fig:pre-processed PT CCDF plots} we represent results for large values of $\delta$ where the distribution function of the waiting times is neither scale-free nor exponential, but, in fact, of Pareto-Tsallis type. This change in the shape of the observed distributions, when compared to the case without pre-processing, shows that the raw data under scrutiny includes periodic components, corresponding to well-defined physical processes, which are however not easily observable through a visual inspection of the data. 
Naturally, these results carry the impact of periodic components of period longer than 29.5 days that have not been accounted for.  In this context, we show in Fig. \ref{fig:pre-processed PT PDF plots} the PDFs (visualised using both linear and logarithmic binning) corresponding to the best identified fit from Fig. \ref{fig:pre-processed PT CCDF plots} to emphasize the idea that CCDF representations alone do not convey the entire information available in the distributions functions. In particular, CCDFs do not capture the fine structure of the PDFs which is due either to real physical processes or are artifacts induced by the small size of the waiting times array, a problem which is particularly relevant at high values of $\delta$.

\section{Conclusion}
\label{sec:conclusion}

We have reported a series of statistical results on the distribution of waiting times for sea level variations in the Port of Trieste using open-source data processing tools and a publicly available dataset that covers more than a century of measurements. Using a definition of waiting times akin to that largely used in econophysics, we find two distinct regimes corresponding to small and large values of the $\delta$ threshold, respectively. In the case of small values of $\delta$ the observed distributions of waiting times are scale-free-like, independent of the pre-processing of the raw data, while for large values of $\delta$ the shape of the distribution depends strongly on how the raw data is processed. Computing the distributions of waiting times using only the raw data yields exponential-like distributions, while a pre-processing that smooths our periodic components up to a period of 29.5 days results in Pareto-Tsallis distributions. 

We expect future studies to focus on exploring to what extent the aforementioned two regimes are generic and can be observed for time series pertaining to research areas as different as space weather and the trading of fiat and cryptocurrencies. Also, we expect software developments on the side of parallel computing with respect to the analysis of waiting times.

\begin{acknowledgement}
For this work, Gabriel Tiberiu Pan\u{a}, PhD student at University of Bucharest, was partly supported through the DANUBIUS-IP European project, Horizon Europe programme, Grant agreement ID: 101079778. The work of Alexandru Nicolin-\.{Z}aczek was supported through the Romanian Ministry of Research, Innovation and Digitalization under Romanian National Core Program LAPLAS VII --- contract no. 30N/2023. The data processing reported here was performed in the computing center of the Faculty of Physics of the University of Bucharest and the data center of the Institute of Space Science -- Subsidiary of INFLPR, M\u{a}gurele, Romania.
\end{acknowledgement}


\begin{thebibliography}{99}
\bibitem{piantadosi2014}Piantadosi, S. Zipf's word frequency law in natural language: A critical review and future directions. {\em Psychonomic Bulletin and Review}. \textbf{21}, 1112-1130 (2014)
\bibitem{lo2013}Lo, C., Bartsch, R., and Ivanov, P. Asymmetry and basic pathways in sleep-stage transitions. {\em Europhysics Letters}. \textbf{102}, 10008 (2013)
\bibitem{lima2017}Lima Predictability of arousal in mouse slow wave sleep by accelerometer data. {\em PLOS ONE}. \textbf{12}, 1-17 (2017)
\bibitem{desousa2020}Sousa, I., Santos Lima, G., Correa, M., Sommer, R., Corso, G., and Bohn, F. Waiting-time statistics in magnetic systems. {\em Scientific Reports}. \textbf{10}, 9692 (2020)
\bibitem{desousa2019}De Sousa, I., Dos Santos Lima, G., Sousa-Lima, R., and Corso, G. Scale-free and characteristic time in urban soundscape. {\em Physica A: Statistical Mechanics And Its Applications}. \textbf{530} pp. 121557 (2019)
\bibitem{kar2020}Kar, A., and Dwivedi, Y. Theory building with big data-driven research – Moving away from the “What” towards the “Why”. {\em International Journal Of Information Management}. \textbf{54} pp. 102205 (2020)
\bibitem{simonsen2002}Simonsen, I., Jensen, M., and Johansen, A. Optimal investment horizons. {\em The European Physical Journal B - Condensed Matter And Complex Systems}. \textbf{27}, 583-586 (2002)
\bibitem{siven2009}Siven, J., and Lins, J. Temporal structure and gain-loss asymmetry for real and artificial stock indices. {\em Phys. Rev. E}. \textbf{80}, 057102 (2009)
\bibitem{pana2021}Pană, G., Ivanoaica, T., Raportaru, M., Băran, V., and Nicolin, A. Towards the Implementation of FAIR Principles on an Earthquake Analysis Platform. {\em 2021 20th RoEduNet Conference: Networking In Education And Research (RoEduNet)}. pp. 1-4 (2021)
\bibitem{vivirschi2020}Vivirschi, B., Boboc, P., Baran, V., and Nicolin, A. Scale-free distributions of waiting times for earthquakes. {\em Physica Scripta}. \textbf{95}, 044011 (2020)
\bibitem{pgt2023}Pană, G., Zgură, S., Băran, V., and Nicolin, A. Waiting times distributions for moonquakes and marsquakes. {\em AIP Conference Proceedings}. \textbf{2843}, 020004 (2023)
\bibitem{pana2023}Pană, G., and Nicolin-Żaczek, A. Motifs in earthquake networks: Romania, Italy, United States of America, and Japan. {\em Physica A: Statistical Mechanics And Its Applications}. \textbf{632} pp. 129301 (2023)
\bibitem{olami1992}Olami, Z., Feder, H., and Christensen, K. Self-organized criticality in a continuous, nonconservative cellular automaton modeling earthquakes. {\em Phys. Rev. Lett.}. \textbf{68}, 1244-1247 (1992)
\bibitem{danubius2019}Friedrich, J., Bold, S., Heininger, P., Bradley, C., Tyler, A., Stanica, A., Bejarano, A., Bellafiore, D., Bowes, M., Brils, J., Brottier, F., Bulla, M., Constantinescu, A., Dabala, C., De Pascalis, F., Ellen, G., Feldbacher, E., Flood, S., Fórizs, I., and Vignati, D. DANUBIUS-RI's Science and Innovation Agenda - International Centre for Advanced Studies on River-Sea Systems.  (2019)
\bibitem{zerbini2017}Zerbini, S., Raicich, F., Prati, C., Bruni, S., Del Conte, S., Errico, M., and Santi, E. Sea-level change in the Northern Mediterranean Sea from long-period tide gauge time series. {\em Earth-Science Reviews}. \textbf{167} pp. 72-87 (2017)
\bibitem{rainich2023}Raicich, F. The sea level time series of Trieste, Molo Sartorio, Italy (1869–2021). {\em Earth System Science Data}. \textbf{15}, 1749-1763 (2023)
\bibitem{seanoe2023}Fabio, R. Sea level observations at Trieste, Molo Sartorio, Italy. SEANOE.. (https://doi.org/10.17882/62758,2023), Accessed: 16-11-2023
\bibitem{polli1947}Polli, S. Analisi periodale delle serie dei livelli marini di Trieste e Venezia. {\em Geofisica Pura E Applicata}. \textbf{10}, 29-40 (1947)
\bibitem{sterneck1905}Sterneck, R.  Kontrolle des Nivellements durch die Fluthmesserangaben und die Schwankungen des Meeresspiegels der Adria. {\em Mittheilungen Der K. U. K. Militar-Geographischen Institutes}. \textbf{XXIV} pp. 75-111, https://anno.onb.ac.at/cgi-content/
\bibitem{panagithub2023}Pană, G. Sea level waiting times toolbox. (https://github.com/gabipana7/waiting-times-sea-level,2023), Accessed: 17-11-2023
\bibitem{gao2023}Gao, T., and Yan, G. Data-driven inference of complex system dynamics: A mini-review. {\em Europhysics Letters}. \textbf{142}, 11001 (2023)
\bibitem{idier2019}Idier, D., Bertin, X., Thompson, P., and Pickering, M. Interactions Between Mean Sea Level, Tide, Surge, Waves and Flooding: Mechanisms and Contributions to Sea Level Variations at the Coast. {\em Surveys In Geophysics}. \textbf{40}, 1603-1630 (2019)
\bibitem{muller2010}Müller, M., Haak, H., Jungclaus, J., Sündermann, J., and Thomas, M. The effect of ocean tides on a climate model simulation. {\em Ocean Modelling}. \textbf{35}, 304-313 (2010)
\bibitem{muis2020}Muis, S., Apecechea, M., Dullaart, J., Lima Rego, J., Madsen, K., Su, J., Yan, K., and Verlaan, M. A High-Resolution Global Dataset of Extreme Sea Levels, Tides, and Storm Surges, Including Future Projections. {\em Frontiers In Marine Science}. \textbf{7} (2020)
\bibitem{clauset09}Clauset, A., Shalizi, C., and Newman, M. Power-Law Distributions in Empirical Data. {\em SIAM Review}. \textbf{51}, 661-703 (2009)
\bibitem{Virkar2014}Virkar, Y., and Clauset, A. Power-law distributions in binned empirical data. {\em The Annals Of Applied Statistics}. \textbf{8}, 89-119 (2014)
\bibitem{alstott14}Alstott, J. powerlaw: A Python Package for Analysis of Heavy-Tailed Distributions (vol 9, e85777, 2014). {\em PLoS ONE}. \textbf{9} (2014,4)
\bibitem{press1992}Press, W., Teukolsky, S., Vetterling, W., and Flannery, B. Numerical Recipes in C (2nd Ed.): The Art of Scientific Computing. (Cambridge University Press,1992)

\end{thebibliography}
\end{document}